# DISEÑO DE SONIDO PARA PRODUCCIONES AUDIOVISUALES E HISTORIAS SONORAS EN EL AULA. HACIA UNA DOCENCIA CREATIVA MEDIANTE EL USO DE HERRAMIENTAS INTELIGENTES.

*(1) Civit, Miguel. Universidad Loyola, departamento de comunicación y educación.*
*mcivit@uloyola.es*

*(2) Cuadrado, Francisco. Universidad Loyola, departamento de comunicación y educación.*
*fjcuadrado@uloyola.es*

## I. INTRODUCCIÓN

El mundo del audio puede resultar muy interesante para gran parte del alumnado, tanto aquellos con inclinaciones creativas como técnicas. La creación y producción musical, su sincronización con imágenes, el desarrollo de podcasts e historias sonoras, el doblaje, etc. son disciplinas que generalmente resultan interesantes pero que pueden tener una barrera de entrada muy elevada debido a su gran complejidad técnica. En ocasiones los no iniciados pueden tardar semanas o incluso meses en empezar a manejar con la soltura necesaria los programas de edición de audio, los cuales no siempre resultan especialmente intuitivos para el alumnado. El aprendizaje mediante el uso de metodologías basadas en problemas (ABP)(Wood,2003) genera, en nuestra experiencia, unos resultados muy superiores a los que pueden observarse mediante el uso de otros métodos docentes como las clases magistrales. Los alumnos adquieren competencias técnicas a la vez que desarrollan proyectos creativos en los que se involucran de manera personal.

A pesar de ello, muchas interacciones entre docentes y alumnado se centran en aspectos de corrección técnica. Estas interacciones consisten, en buena parte, en aspectos que, para el profesional del tema, son prácticamente automáticos pero que son difíciles para alumnos principiantes. Estos incluyen desde configurar distintos parámetros en reverbs (pre-delay, decay...) a cómo limpiar diálogos en mal estado, etc.; la cantidad de herramientas con las que trabajar el audio es increíblemente extensa y muchas de sus características pueden presentar diferencias significativas dependiendo del programa utilizado. Estos problemas no hacen sino acrecentarse cuando el alumno se enfrenta al reto de componer o modificar música que ha de ir en sintonía con la imagen o como acompañamiento a un podcast o historia sonora, especialmente si carecen de un cierto bagaje en el ámbito de la composición.

El uso de herramientas basadas en inteligencia artificial para la educación en diseño de sonido (HIAEDS), empleada en distintos roles dentro del aula, puede disminuir la barrera de entrada a la materia, así como incrementar, de forma significativa, la autonomía en el aprendizaje de los estudiantes (Ge, 2018). Basándonos en la clasificación que hace Avdeeff (2019) de las herramientas para la producción de materiales sonoros (música en su caso), las herramientas inteligentes pueden no solamente ser usadas para la mejora técnica del producto final, sino también en muchos casos para la agilización de los procesos más técnicos y tediosos de los procesos creativos. De esta manera, no solamente podemos facilitar el aprendizaje a nuestro alumnado mediante el uso de herramientas inteligentes, sino también crear un espacio dónde la creatividad y la narrativa de los proyectos se vea facilitada y potenciada gracias al uso de las mismas.

En el curso 2021/2022 en la asignatura de impartición en inglés Digital Sound Design (4º de Informática de las tecnologías virtuales, Universidad Loyola, Sevilla) probamos varias herramientas basadas en IA tanto centrándonos en la docencia como, fundamentalmente, en la creación de contenido sonoro. La integración de estas herramientas en la continuidad de la asignatura partió tanto de los docentes como de la propia iniciativa del alumnado. Así mismo, estamos aplicando las metodologías y herramientas que han resultado más exitosas con la nueva promoción 2022/2023. Esta experiencia docente servirá de base para la discusión sobre las aplicaciones de dichas herramientas en este congreso.

## II. OBJETIVOS

Nuestro objetivo es generar una discusión sobre la aplicación de herramientas inteligentes en la docencia del diseño de sonido. Todo ello como vía para reducir la barrera de entrada inicial que plantea el diseño de sonido y crear espacio para un aprendizaje basado en la creatividad en lugar de en los detalles tecnológicos más farragosos. Debido a la reducida muestra de alumnos con la que se cuenta, es nuestro objetivo futuro generar un estudio cuantitativo, que pueda aportar una mayor cantidad de elementos analizados, a lo largo de al menos cuatro promociones de estudiantes. Es por ello por lo que el estudio actual, con una promoción completa y otra en curso, centra sus objetivos en la discusión de las posibilidades de estas herramientas para el diseño de sonido con finalidades docentes, adaptándolas a la clasificación de Lukin, (2016).

La integración final de estos sistemas expertos en el aula es el objetivo final de este proyecto y por ello crearemos un modelo genérico para implementar estas distintas herramientas en el ABP basándonos en los distintos tipos de las mismas y la posibilidad de implementarlas en las distintas etapas del ciclo de aprendizaje experimental (Kolb,2014).

### III. METODOLOGÍA

En este trabajo de investigación discutiremos algunas de las implicaciones que las herramientas basadas en IA tienen en la docencia del diseño de sonido y las adaptaremos a una clasificación preestablecida. Basándonos en un estudio experiencial y cualitativo de la impartición de la asignatura de Digital Sound Design, a la promoción 21/22 en la Universidad Loyola, analizaremos los resultados obtenidos por los alumnos que han empleado estas nuevas tecnologías y compararemos sus resultados con los de los compañeros que han realizado las mismas actividades académicas sin el uso de IA.

Debido a la necesidad de generar datos específicos con los que poder evaluar de la manera más objetiva posible el impacto real de estas herramientas utilizaremos un método cuantitativo, adaptado del estudio realizado por Corral (2019) que también centraba su atención en la enseñanza técnica a usuarios no expertos. Dichos resultados los comentaremos brevemente ya que, debido a la necesidad de una muestra más amplia para obtener resultados estadísticamente significativos, formarán parte de un estudio posterior que analice el impacto del uso de herramientas basadas en AI frente al uso de otras basadas en ejemplos previamente construidos a lo largo de entre cuatro y seis promociones de estudiantes.

Dicho método recoge datos cuantitativos mediante el uso de cuestionarios estandarizados para los diferentes tipos de proyectos audiovisuales realizados durante el curso. Debido a que todos los proyectos se elaboran en grupos de 3 a 5 estudiantes formados aleatoriamente, los resultados se tomarán como la media de las puntuaciones o mediciones entre los distintos estudiantes de cada grupo. De esta manera podemos organizar los datos por estudiantes, grupos y proyectos concretos (cuyas características particulares pueden afectar algunas de las variables) y dividirlos entre aquellos que han usado herramientas inteligentes y aquellos que no.

Los datos recogidos mediante los cuestionarios estandarizados son:

- Percepción subjetiva de la dificultad técnica del proyecto ( del 1 al 4) (DT)
- Percepción subjetiva del interés por el proyecto (del 1 al 4) (IP)
- Percepción subjetiva del riesgo creativo asumible en el proyecto teniendo en cuenta la cantidad de tiempo total para finalizarlo (del 1 al 4) (RC)
- Percepción subjetiva de la calidad del resultado final (1 al 4) (CF)

Así mismo se miden los siguientes datos:

- Tiempo medio empleado planeando el proyecto (contando períodos de escritura de guiones, división del trabajo, etc.) por alumno (en minutos) (TP)
- Tiempo medio empleado utilizando herramientas de edición o producción de audio (inteligentes o no) por alumno (en minutos) (TE)
- Nota conseguida en el proyecto (del 1 al 10)

La obtención y procesado de estos datos nos permitirá tener una idea general de la percepción de competencia de los estudiantes, su predisposición a ser creativos y asumir riesgos, así como del resultado general obtenido en los proyectos. Todo ello permite una comparativa necesaria sobre el uso de las HIAEDS en el aula frente a la realización de las actividades sin el empleo de las mismas.

Para la definición formal del estudio se ha empleado la metodología propuesta en Crawford (2014). El diseño formal del estudio a realizar se resume en la Ilustración 1.

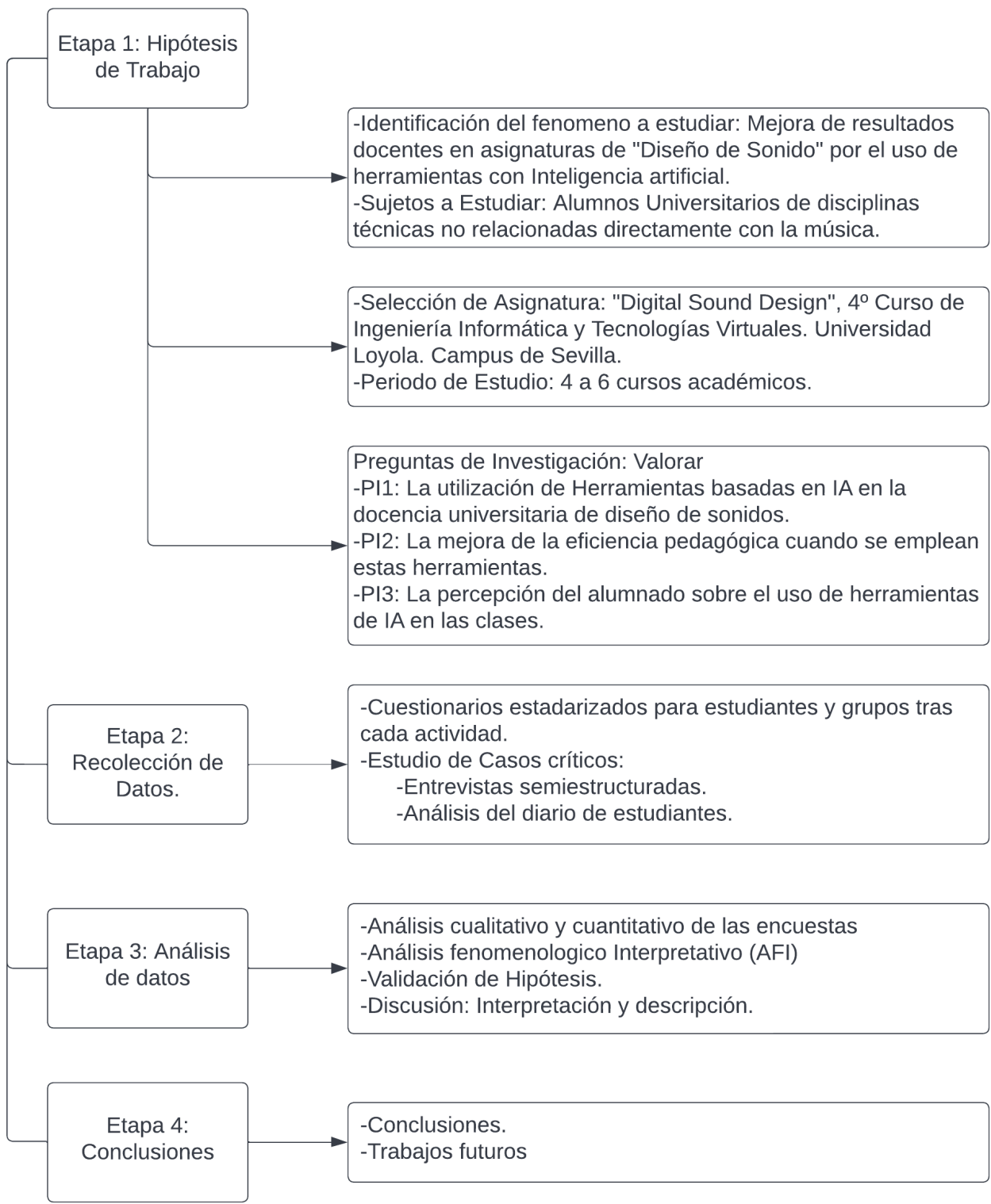

*Ilustración 1: Diseño formal del estudio.*

## IV. CLASIFICACIONES DE HERRAMIENTAS INTELIGENTES PARA EL DISEÑO DE SONIDO.

En 2016 Rose Luckin publicó un interesante artículo sobre el futuro de la educación gracias a las herramientas basadas en IA (AIED). Si bien las herramientas basadas en IA para el audio constituyen un campo en el que la evolución es increíblemente rápida y con un increíble desarrollo desde 2017 (como podemos observar en Civit M. (2022) con herramientas para generación automática de música), es cierto que el artículo plantea una clasificación general de las posibles herramientas para la docencia que puede resultar muy útil dada la variedad de situaciones que refleja. Varios de los apartados de la clasificación se encuentran aún en un estado de desarrollo muy temprano en lo que respecta a la viabilidad de las herramientas existentes en el campo de la enseñanza de diseño de sonido, pero es precisamente por ello que resulta especialmente relevante su adaptación a dicho ámbito.

A continuación, exploraremos dicha clasificación adaptándola al campo del diseño de sonido, proponiendo algunas herramientas o soluciones concretas que pueden implementarse actualmente y aclarando cuáles categorías han sido empleadas con éxito en nuestra experiencia docente. Estas se agrupan en las siguientes tres grandes familias:

i) Docencia personalizada.

ii) Soporte inteligente para el aprendizaje colaborativo

iii) Realidad virtual inteligente.

A pesar de la relevancia genérica de la clasificación, en el campo del diseño de sonido, y hasta cierto punto en el de las artes audiovisuales, podemos perfilar de manera más específica cada una de las categorías.

Cuando nos referimos a docencia personalizada basada en IA, Luckin comenta como el uso de las IA puede generar herramientas que se encarguen de tutorizar de manera individual a cada estudiante atendiendo a sus características y necesidades específicas, corrigiendo sus fallos concretos y ampliando materiales conforme a sus intereses y capacidades. Actualmente desconocemos que existan herramientas que aporten este nivel de detalle y de individualización en el campo del diseño de sonido con la excepción (si lo consideramos parte de este campo) de la enseñanza musical. Herramientas como los juguetes musicales de Google (Cuadrado, 2019) pueden aportar en casos concretos desafíos musicales que se adapten al usuario en términos de complejidad. Así mismo existen algunas herramientas, como señala Calderon-Garrido (2021), para la enseñanza de instrumentos musicales que pueden elegir cambiar las lecciones para intentar corregir los fallos técnicos que demuestre el intérprete. Otros campos del diseño de sonido como la mezcla o el mastering de audio, disponen de herramientas capaces de resolver problemas reales mediante inteligencia artificial resultado más que aceptable. Herramientas como Ozone de la empresa iZotope(Collins,2021), son capaces de masterizar canciones, podcasts o películas concretas. Los estudiantes pueden observar los cambios concretos que realizan estos plugins (herramientas de edición de audio) y aprender de los mismos mediante imitación y a través de descubrir por ellos mismos las razones por las que la IA ha ejecutado ciertos cambios en el audio. Este sistema supone una clara ventaja sobre el aprendizaje mediante el uso de "presets" genéricos en plugins de edición de audio, pues en vez de aprender copiando una serie de configuraciones que funcionan para muchos casos genéricos, los estudiantes pueden presentarle a la IA infinidad de problemas para los que obtienen configuraciones específicas de las que poder extrapolar conclusiones y normas generales. Un área pendiente de desarrollo en el caso de las AIED es el de las herramientas inteligentes explicables (Khosravi, 2022), a pesar de que las herramientas nombradas muestran los distintos parámetros que modifican para obtener un resultado final, éstas no explican el porqué de las decisiones que toman de manera que queda en última instancia en manos del instructor clarificar si las conclusiones a las que llega el alumnado son correctas.

La segunda rama de la clasificación corresponde a aquellos sistemas que facilitan la colaboración de los estudiantes en proyectos, actividades o aprendizaje en grupo. Esta clasificación se subdivide a su vez en:

- Formación adaptativa de grupos: Basada en IA, la formación de los grupos puede adaptarse a las necesidades de los estudiantes y formar grupos que aprovechen las distintas habilidades de estos o que fomenten el aprendizaje de un recurso específico.

- Facilitación experta: Creando plataformas o entidades capaces de cambiar dinámicamente las actividades o de facilitar consejos en puntos cruciales del aprendizaje.

- Agentes virtuales inteligentes: Las IA pueden formar parte de los grupos de estudiantes bien como expertos (similar al uso antes mencionado de aplicaciones capaces de resolver problemas concretos), como otro compañero o como alguien a quien enseñar.

- Moderación inteligente: Fomentando el debate y moderándolo, pudiendo intervenir para guiarlo en direcciones concretas.

Esta área de las inteligencias artificiales se encuentra muy en su infancia en el campo del diseño de sonido con algunas excepciones. En el caso de la formación adaptativa de grupos y de la moderación inteligente, pueden usarse AIED de carácter general, pero las mismas pueden no tener en cuenta las características específicas del diseño de sonido. Por otra parte, el uso de agentes virtuales externos es un proceso en desarrollo, que requiere a empresas como iZotope que desarrollen inteligencias explicables con tooltips que permitan a los estudiantes, no sólo entender los motivos detrás de las decisiones tomadas, sino también corregir errores que estos puedan cometer durante los procesos de edición de audio (reconocimiento de clics, peaking, saturación, problemas de fase, descompensación de distintas bandas de frecuencia, etc.) y expliquen cómo solucionarlos. Por último, en el campo de la educación musical si existen plataformas digitales como SmartMusic de Finale (Griffin, 2008) con la capacidad de integrar numerosas herramientas para el aprendizaje musical y con la capacidad intrínseca de colaboración entre distintos estudiantes y profesores. A pesar de ello, sus capacidades de autocorrección, aprendizaje automático y guía basadas en IA pueden desarrollarse en mayor profundidad Wei (2022). Otra área donde el desarrollo es más que necesario es en la de la creación de estas plataformas colaborativas con herramientas en AIED para edición y procesado de audio. Empresas como Steinberg, desarrolladora de la DAW (Digital Audio Workstation) Cubase o Apple, con su DAW Logic Pro y con más que suficiente capacidad de desarrollo de herramientas AI, (Podolny, 2020) es probable que ocupen este nicho de mercado con plataformas similares a SmartMusic pero diseñadas para la producción sonora. Algunas de estas plataformas para la colaboración en la producción sonora han sido creadas, especialmente durante la pandemia de Covid-19 para paliar las consecuencias del aislamiento, pero su mayor foco de atención es en la baja latencia, su compensación y la conversión a formato remoto de herramientas ya disponibles, más que en herramientas de ayuda a los usuarios o diseñadas para la educación. (Hoene,2021).

La tercera categoría, realidad virtual inteligente, puede ser muy útil para distintos tipos de aprendizaje y es sin duda un tema de gran interés actual con grandes figuras como Meta o Apple trabajando en ella. En nuestro campo de estudio específico sería interesante reconsiderar esta categoría en la actualidad hasta que existan más implementaciones prácticas. Si bien es cierto que la creación musical y los conciertos virtuales son eventos recurrentes de éxito, como puede observarse en el concierto de Travis Scott en el videojuego Fortnite con 12.3 millones de asistentes simultáneos (Clement, 2021), las colaboraciones para producción y diseño de sonido en ambientes virtuales es otro campo con gran potencial debido a la inmersión posible y las herramientas disponibles. Herramientas basadas en realidad virtual o aumentada son cada vez más frecuentes en nuestro campo, como pueden ser controladores midi mediante el uso de dispositivos de detección de movimiento como Leap Motion (Silva, 2013) con la capacidad de representación de objetos en 3D mediante realidad aumentada basada en IA y feedback háptico sonoro, o la realización de mezclas con audio espacial en Dobly Atmos mediante el uso de realidad virtual (pudiendo visualizar con las gafas RV los focos de los distintos elementos sonoros de la mezcla).

Un grupo de herramientas basadas en IA que son de especial relevancia para el diseño de sonido y que podríamos incluir en una versión ampliada de esta categoría son las herramientas que permiten crear medios audiovisuales complementarios a nuestras creaciones sonoras. Mediante el uso de herramientas generadoras de imágenes como DALLE o Midjourney (Garrido-Merchán, 2022) estudiantes sin formación artística y sin medios para colaborar con estudiantes o profesionales de disciplinas artísticas pueden crear desde imágenes como portadas de historias sonoras, a novelas gráficas sonorizadas, videoclips musicales o diseños gráficos para personajes animados en videojuegos. Esta habilidad les permite implicarse a nivel creativo y motivacional más profundamente con los proyectos que pueden desarrollar en clase, dónde muchas veces el diseño de sonido está supeditado a algún tipo de contenido audiovisual.

Podríamos entonces volver a imaginar la clasificación de Lurkin (2016) adaptándola al diseño de sonido:

I. Docencia personalizada y sistemas expertos de procesado de audio (Bocko, 2010)(Moffat,2021) dentro de este ámbito podemos encontrar sistemas expertos de producción y mezcla. También entrarían en este apartado las herramientas de composición automática o de ayuda para completar composiciones parciales, Como ejemplo Magenta Studio (Roberts,2019) es un plugin para Ableton que nos permite generar, ampliar e interpolar melodías y también crear pistas de percusión o humanizar pistas existentes mediante Inteligencia Artificial. Logic Pro también incorpora una herramienta para la generación de percusión basada en IA (Vogl,2017).

II. Soportes inteligentes para el aprendizaje colaborativo. Los sistemas de aprendizaje colaborativo cada vez se están empleando más en el ámbito de la educación superior. Tal como nos indica Montebello (2018) entre sus ventajas tenemos beneficios en los ámbitos sociales, psicológicos, académicos y de evaluación. Una aplicación muy interesante al campo de la música se encuentra en

de Bruin (2022).

III. Realidad virtual inteligente: Las tecnologías de realidad virtual inteligente están siendo aplicadas al diseño de sonido para películas (Zhang,2022). Tanto la realidad virtual como la realidad aumentada tienen un importante potencial dentro de la educación musical (Yeon, 2018) en general y de la educación en tecnología musical (Cook,2019) en particular.

IV. Generadores de contenido multimedia inteligentes: este tipo de generador son enormemente útiles en muchos casos. Bell (2020) estudia, por ejemplo, su aplicación en el caso de la educación musical de estudiantes con discapacidad.

# V. APLICACIÓN DE HIAEDS A PROCESOS DE ABP

En todo proceso de aprendizaje, y el diseño de sonido no es una excepción, la práctica reflexiva es fundamental, puesto que capacita a la persona para aprender de sus propias experiencias. En nuestro caso, a través del aprendizaje basado en problemas, pretendemos que las experiencias obtenidas de la realización de proyectos constituyan el núcleo central del proceso de aprendizaje. Ya en los años 80 del siglo pasado Kolb (2014) desarrolló una teoría del aprendizaje experiencial, que ha sido muy ampliamente empleada desde entonces en el ámbito de la enseñanza musical entre otros (Russell-Bowie, D., 2013), a la que denomino Ciclo de aprendizaje (Ilustración 2). El ciclo de Kolb está formado por cuatro fases que nos permiten ingresar en cualquier punto, pero todas las etapas deben seguirse en secuencia para lograr un proceso de aprendizaje exitoso. Las fases del ciclo son: Experiencia Concreta, Observación Reflexiva, Conceptualización Abstracta y Experimentación Activa. En el caso del diseño de sonido muchas veces empezamos con una experimentación activa, planificando el proyecto concreto y probando distintos parámetros para "ver" como suenan y como responde la tecnología, tras tomar alguna decisión obtendremos una experiencia concreta que tras un periodo de revisión nos llevará a juicios de valor sobre los resultados y a una conceptualización abstracta sobre posibles mejoras, motivos por los que distintos elementos suenan de una manera específica, etc. Ello nos llevará de nuevo a un proceso de experimentación activa dónde tendremos que probar nuevos plugins, configuraciones o ediciones en función de las conclusiones obtenidas previamente. Este proceso lo hacemos de forma iterativa de modo que al experimentar producimos experiencias concretas que nos llevan a nuevas reflexiones y ha afianzar, con ayuda de ella, nuestros conocimientos abstractos. Es importante hacer notar que el uso de algunas herramientas basadas en IA nos ayuda a reducir, en buena medida, muchos detalles concretos de la experimentación activa permitiendo que ésta pueda en un proceso más creativo y fructífero y ayudando, de este modo, a permitir dedicar más tiempo a la reflexión creativa y a la adquisición de conocimientos abstractos.

2)

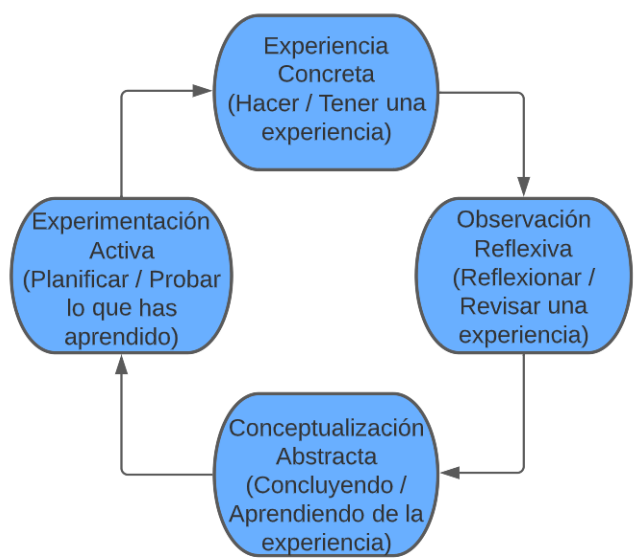

*Ilustración 2: Ciclo de Aprendizaje.*

Al intentar resolver un problema concreto dentro de uno de los proyectos de procesado sonoro, el alumno identifica los recursos de técnicos específicos que necesita (ecualizadores, compresores,…) y establece unos criterios y evidencias que le permitirán conocer la corrección de su solución al ejercicio propuesto (Experimentación Activa). Luego, configura los diversos plugins y experimenta una posible solución y, si no hay errores en la configuración empleada, el resultado sonoro reproducido en la DAW coincidirá con sus expectativas o con sus pistas de referencia (Acción). Cuando el alumno escucha una discrepancia significativa entre lo que se reproduce y sus expectativas o referencias, éste comprende que existe un problema con el plugin o configuración implementada y reflexiona sobre lo que ha sucedido (Observación Reflexiva).

Finalmente, el alumno extrae conclusiones del análisis de la información disponible para aprender de su experiencia (Conceptualización abstracta). Utiliza los criterios y evidencias previamente establecidos para determinar el grado de consecución de su objetivo -una correcta programación del sonido a generar con una correcta solución- o no -la falta de coherencia del sonido generado con los objetivos preestablecidos. En este último caso, el estudiante utilizará sus conclusiones para pasar a una etapa posterior de Experimentación Activa para planificar una estrategia revisada (es decir, hacer las correcciones correspondientes y, si no hay errores, generar de nuevo los sonidos). De este modo continuará lograr su objetivo haciendo que el ciclo de aprendizaje comience de nuevo.

Así, el ciclo de Kolb implica que no basta con tener una experiencia para aprender, sino que debe haber un proceso de reflexión posterior a la experiencia: es fundamental reflexionar sobre la experiencia para formular conceptos que puedan aplicarse a nuevos entornos. Finalmente, este nuevo aprendizaje se pondrá a prueba en nuevas situaciones y contextos. De esta forma, teoría, acción, reflexión y práctica se vinculan en un ciclo dinámico y se complementan (Shon, 1991).

Finalmente, con la ayuda del soporte tecnológico que brindan las herramientas basadas en inteligencia artificial, se espera incentivar al alumno a aplicar cualidades más creativas en el desarrollo de sus proyectos. En la siguiente sección VI – Una experiencia docente, comentaremos más en detalle como implementamos estas teorías y dinámicas en la elaboración de la programación de una asignatura de diseño de sonido.

## VI. UNA EXPERIENCIA DOCENTE

Durante el curso 2021/2022 impartimos la asignatura Digital Sound Design (diseño de sonido digital) a alumnos de 4º de Ingeniería Informática y Tecnologías Virtuales de la Universidad Loyola Andalucía. El curso se organizó aprovechando las metodologías Flipped Classroom, aprendizaje informal y fundamentalmente centrándonos en el aprendizaje basado en proyectos. Con la ayuda de estas metodologías docentes se creó un ambiente distendido en las clases donde gran parte del conocimiento es investigado y adquirido por los propios alumnos, siendo la labor fundamental del docente resolver problemas complejos y dudas, organizar el contenido y orientar al alumnado en la realización del proyecto. Así, las clases parecían más equipos creativos "profesionales" supervisados que clases magistrales dónde se "manda tarea" sobre los temas tratados.

La asignatura consta de 6 créditos ECTS, dura un semestre y se organiza en torno a tres grandes proyectos de creciente complejidad:

- Realización de una historia sonora: Requiere todos los pasos desde la escritura del guion, la grabación de voces y efectos de sonido y el procesado y edición de música de bibliotecas para adaptarlo a la historia. Al ser un proyecto puramente basado en audio los alumnos pueden centrarse en crear su primer proyecto de diseño de sonido y en familiarizarse con las distintas herramientas de edición y procesado de audio.

- Sonorización de una escena: Requiere sonorizar con doblaje, música y efectos de sonido una escena de corta duración escogida por los alumnos con la aprobación del profesorado. Apoyándose en los conocimientos adquiridos en el proyecto anterior, éste tiene una capa más de dificultad al incluir la sincronización con imágenes como un elemento adicional. La creatividad vuelve a ser central en el proyecto ya que los alumnos pueden reescribir diálogos y cambiar el sentido completo de la escena a través de su sonorización (lo que en esta promoción llevó a la realización de parodias y versiones satíricas).

- Diseño de sonido para un videojuego creado por los propios estudiantes en la asignatura paralela Narrativa Multimedia: Requiere que los estudiantes se familiaricen con el audio adaptativo, junto con la creación de numerosos efectos de sonido con múltiples permutaciones de los mismos.

Debido al propio perfil de los estudiantes (informáticos en su 4º curso) y al tema central de la tesis de uno de los profesores (generación de música con inteligencia artificial), la idea de usar herramientas basadas en IA para el diseño de sonido surgió espontáneamente en clase entre docentes y alumnado. Por ello, las aplicaciones de dichas herramientas son muy variadas en esta primera promoción, sirviendo en parte como filtro para la elección de herramientas en futuras promociones.

Algunas de los usos que se dieron de las HIAEDS en los distintos proyectos fueron:

- Transcripción de canciones mediante AI: Gracias a ello, estudiantes con conocimiento bastante limitados en armonía y composición pudieron transcribir secuencias de acordes de canciones famosas que les sirviesen de base para realizar "covers" que emplear como banda sonora de videojuegos.
- Conversión de audio a midi: A pesar de que este proceso puede realizarse con otros tipos de tecnología, la conversión permitía a los estudiantes con conocimientos sobre composición melódica limitados transcribir melodías pegadizas y reutilizarlas con instrumento de distintos timbres.
- Aislamiento de sonidos de su entorno: Usado tanto para reducir ruido en grabaciones hechas en entornos subóptimos (para los proyectos de sonorización de escena) como para la reutilización de patrones de baterías como simples.
- Aprendizaje mediante imitación de sistemas expertos y resolución de problemas complejos: Como indicamos en la sección IV- clasificaciones de herramientas inteligentes para el diseño de sonido, los alumnos fueron capaces de resolver algunas situaciones de mezcla de voces, efectos de sonido y música complejas mediante el uso de sistemas expertos de los que pudieron aprender y modificar parámetros hasta obtener resultados satisfactorios.
- Generación automática de música para BSO y creación automática de arte de referencia para el diseño de videojuegos.

En observaciones de los docentes, el uso de estas herramientas no supuso necesariamente un resultado final mejor de los grupos que las emplearon sobre los que no. En general el nivel de la promoción fue alto y resultados finales similares, o en ocasiones mejores, pueden obtenerse mediante el uso de otras tecnologías más consolidadas y que generalmente funcionan con mayor fiabilidad. El factor diferenciador fundamental del uso de estas tecnologías es, en este caso, su capacidad para permitir que los estudiantes exploren actividades creativas como la composición con resultados más que aceptables, desarrollando sus capacidad para innovar y crear productos propios con los que se sienten profundamente identificados frente a aquellos que emplearon materiales ajenos "descargados de librerías stock" u otras fuentes, cuyos trabajos podían carecer de ese factor creativo e innovador. Por otra parte el uso de algunos de estos sistemas, también permitió al alumnado resolver problemas de manera autónoma sin tener que asistir a la ayuda de los profesores o sin tener que optar por abandonar ciertas ideas al convertirse en demasiado complejas.

### VII. CONCLUSIONES

A pesar de que, como comentamos en más profundidad en el apartado III – Metodología, la experiencia docente queda aquí reflejada mediante métodos cualitativos, los resultados en cuanto a viabilidad, para su uso en las aulas de diseño de sonido de las herramientas basadas en IA, parece clara. Queda reflejado en el estudio como mediante el uso de dichas herramientas los estudiantes asumen mayores riesgos, lo que les lleva a desarrollar tanto su capacidad creativa como competencias que a priori les pueden parecer inasumibles.

El estudio de los resultados finales de los proyectos desarrollados en relación con el uso de las HIAEDS es claramente necesario, pues los datos obtenidos hasta la fecha pertenecen a una muestra demasiado pequeña para resultar matemáticamente significativos. Además, estos datos nos muestran información inconcluyente en lo que respecta a la calidad técnica final de los productos audiovisuales, y es por ello un campo interesante en el que seguir investigando para establecer si, conforme avancen estas nuevas tecnologías, los resultados finales son significativamente mejores gracias a su uso, con independencia de las competencias que promuevan durante el desarrollo de los estudiantes. A pesar de ello, creemos que la metodología expuesta puede ser útil no sólo en el estudio futuro HIAEDS sino también en otros estudios sobre implementación diversas tecnologías basadas en IA en la enseñanza de diferentes materias audiovisuales.

De la misma manera, las clasificaciones empleadas, adaptadas al campo de la enseñanza en diseño de sonido

(y extensibles a otras materias relacionadas con medios audiovisuales) pueden servir de orientación no sólo a educadores, sino también a aquellos desarrollares de Software con intenciones de desarrollar o mejorar las tecnologías de IA con objetivos pedagógicos, tanto en el campo de la enseñanza superior, como en el del aprendizaje autónomo y la colaboración entre profesionales.

**Bibliografía**

# VIII. Agradecimientos

.